\pdfoutput=1
\documentclass[letter,12pt]{article}
\usepackage{fullpage}
\usepackage{natbib}
\usepackage{authblk}
\usepackage{times}
\usepackage{latexsym}
\usepackage[T1]{fontenc}
\usepackage[utf8]{inputenc}
\usepackage{microtype}
\usepackage{inconsolata}
\usepackage{enumitem}
\usepackage{pgfplots}
\usepackage{algorithm}
\usepackage{algpseudocode}
\usepackage{graphicx}
\usepackage{graphics}
\usepackage{placeins}
\usepackage{tabularx}
\usepackage{makecell}
\usepackage{multicol}
\usepackage{booktabs} 
\usepackage{array}  
\usepackage{dblfloatfix}
\usepackage{verbatim}
\usepackage{colortbl}
\newcolumntype{P}[1]{>{\centering\arraybackslash}p{#1}}
\newcolumntype{M}[1]{>{\centering\arraybackslash}m{#1}}

\usepackage{subcaption}
\usepackage{amssymb}
\usepackage{amsmath}
\usepackage{hyperref}
\usepackage{mathtools}
\usepackage[normalem]{ulem}

\definecolor{red}{rgb}{1.0, 0.29, 0.33}

\def\1{\bm{1}}
\definecolor{bblue}{HTML}{4F81BD}
\definecolor{rred}{HTML}{C0504D}
\definecolor{ggreen}{HTML}{9BBB59}
\definecolor{ppurple}{HTML}{9F4C7C}
\definecolor{oorange}{HTML}{FFA500}

\pgfplotsset{width=8cm,compat=1.9}
\graphicspath{ {./media/} }

\title{Overview of the TREC 2025 Product Search and Recommendation Track}
\author[1]{Dean E. Alvarez\thanks{Corresponding author: deana3@illinois.edu}}
\author[2]{Surya Kallumadi}
\author[3]{Daniel Campos}
\author[1]{ChengXiang Zhai}
\author[4]{Alessandro Magnani}
\author[5]{Rikiya Takehi}
\author[6]{Michael D. Ekstrand}
\affil[1]{Department of Computer Science, University of Illinois Urbana-Champaign}
\affil[2]{Coursera}
\affil[3]{Snowflake}

\affil[4]{Coupang}
\affil[5]{Waseda University}
\affil[6]{Department of Information Science, Drexel University}

\date{}
\begin{document}
\maketitle

\begin{abstract}
In the past few years, consumers have moved the bulk of their product exploration and purchasing efforts online seeking speed, convenience, and price comparison with ease unimaginable for in-person shopping. As product catalogs have grown in diversity and size product search and recommendation have become a cornerstone for e-commerce sites. 

Despite the widespread usage of search engines in e-commerce, there is no high-quality dataset designed to evaluate end-to-end retrieval quality. In 2025, we ran a revised and continued version of the Product Search track previously run at TREC 2023 and TREC 2024.
The 2025 product search track had two tasks: \textbf{query expansion} and \textbf{related-product recommendation}.
The related-product recommendation task is particularly novel, providing an annotated data set of product relationships that distinguishes between complementary and related products.  We anticipate the data from this track will enable better recommendation and search applications that reflect user needs, as a building block for conversational product discovery experiences.
\end{abstract}

\section{Introduction}

This paper provides an overview of the 2025 TREC Track on Product Search and Recommendation. This marks the third year of the Product Search and Recommendation Track. 
This year, based on the experience of the past years, we focused on two tasks: the \textbf{search task} and the \textbf{recommendation task}. 

\paragraph{Search Task.}
For the last few decades, TREC has led the Information Retrieval (IR) research community by assembling tracks around areas of growing interest and importance. We believe product search is a novel and growing area which deserves further study both in terms of ranking and evaluation methods. As consumers around the world are continually shifting their purchasing online all companies that sell goods become e-commerce companies and product catalogues become intractable to browse. From research perspective, product search raises many interesting new challenges and has attracted much attention recently. For example, a SIGIR Workshop on eCommerce has been organized annually since 2017~\cite{kale2022ecom}. We have also seen increasingly more research publications on product search.   

In the past two years, we had mainly had two separate tasks, text-based retrieval and multi-modal retrieval. Each of these tasks was powered by large datasets featuring millions of items and tens of thousands of training queries and samples. While this collection already featured a wealth of evaluation labels, this evaluation was not performed with the rigor associated with TREC nor was labeling done with the goal of evaluating end-to-end retrieval. While there are a few datasets which seek to study product search, such datasets are small and focused on re-ranking retrieval sets which have already been generated by large e-commerce companies \citep{wands, reddy2022shopping}. As there is no end-to-end dataset, there is no current benchmark for evaluating end-to-end product retrieval. By running the TREC 2023 and 2024 Product Search Track, we were hoping to fill this unmet need with a reusable test collection which can illuminate many interesting new research topics. For example, we could make a comparison of how systems from other domains (e.g. open-domain web search) transfer to product search, how readily (e.g. is zero-shot transfer possible), and what challenges arise. This year, we expand this line by focusing on query reformulation. The search task is detailed in section \ref{sec:search}.

\paragraph{Recommendation Task.}
In addition to the product search task, we added a recommendation task. Recommendation has been a very large part of the IR research community, yet no previous TREC Tracks have particularly focused on this yet-enlarging field. The community of RecSys is always in lack of \textit{high-quality} datasets, and we believe a TREC-quality recommendation dataset would have immense possibility in the growth of the field. 

Particularly, we focus on item-to-item relationships. Nuanced information of item-to-item relationships is usually very difficult to obtain, as we could not know this from implicit feedback. Consequently, this has restricted the researchers and industry platforms to test and improve algorithms for various tasks. For example, complementary product recommendation could be a task that could be improved from having nuanced information of item relations. Although complementary product recommendation (or more generally item-to-item recommendation) has been a very large field of research and is very relevant to industry applications, implicit feedback datasets like co-purchases were often used to evaluate such systems~\citep{zijie2020nextitem, xu2020complementknowledge}. These datasets contain the user's ratings, possibly an interaction log, and may contain information from the source platform such as its own recommendation outputs (e.g. Amazon ``related products''). The most widely-used product data sets for item-relations is from Amazon \citep{ni2019amazonreview}. This dataset includes user reviews to Amazon products, which also has a list of often co-purchased products and often co-viewed products. However, these are implicit information that are collected by an unknown \textit{internal} algorithm. We know from various work that implicit datasets are often severely biased and often mistreating some products~\citep{yao2018judgingsimilarityrelateditem, tobias2020itemtoitem}. We design a high-quality dataset with manually annotated item relations, helping the community to design and evaluate such algorithm. The recommendation task is detailed in section \ref{sec:recommendation}.
\section{Product Dataset Overview}
\label{sec:data-over}
Our product dataset in 2023 and 2024 were novel retrieval datasets built by processing the 2022 KDD Cup ESCI Improving Product Search (ESCI) dataset \cite{reddy2022shopping} \footnote{https://www.aicrowd.com/challenges/esci-challenge-for-improving-product-search?source=mlcontests} similar to how the MS-MARCO Passage Ranking Dataset was processed \cite{nguyen2016ms}. The ESCI dataset features 33,804 queries sampled from Amazon's search logs in English, Spanish and Japanese. To generate the collection we leveraged the product index from the KDD Cup 2022 product substitute classification challenge and reformatted it to match that of a TREC collection. Then, for each product, we have extracted and processed relevant product images and metadata which did not exist in the public domain previously. Overall, the collection features 1,803,063 products with related metadata such as brand, color, title, and description.

To aid rapid iteration we have created triplet files for each query featuring positive examples and negative samples generated using BM25, hard negatives, and products labeled substitute or complementary. 

In the TREC setup:
\begin{itemize}
    \item Document corpus: The 1.8 million products from the ESCI Datasets. In CatA runs, retrieval is from the full corpus while in CatB, retrieval is on a provided top 1000 BM25 retrieved subset for reranking.
    \item Queries: We provide 30,000 training queries and 3,000 development queries which feature labeling of an average of 23 product item pairs per query. There are an additional 1,000 queries which do not have any publicly released relevance judgments which we will use to extract to a small amount of queries for TREC evaluation. We used LLMs to generate queries that constitute the test collection.
    \item Judgments: Each query in the dataset has at least 1 item in the collection which is relevant and for many queries there are multiple relevant items. We hope that we can use pooled judging runs to get a more comprehensive view of product relevance.
\end{itemize}
For evaluation metrics, we measure recall at various depths using the ESCI annotation and will measure MAP, NDCG, and RBP using pooled TREC judgments. No special assessment protocols are proposed for evaluation as we 
hope to first use the same evaluation methods used for Web search so that we can directly compare the performances of retrieval algorithms on product search with those on Web search.
\section{Search Task}\label{sec:search}
Search engines generally work well when a user’s query is effective, i.e., when the query uses the “right” keywords that are matched in relevant documents but not in non-relevant documents. Indeed, the ideal query hypothesis states that for any information need, there exists an ideal keyword query that would be so effective to ensure that a search engine would return at least one relevant document and rank it at the very top. For example, if we take a whole relevant document as a query, presumably the same document would be ranked on the top. The ideal query hypothesis suggests that if we can formulate an effective query, the traditional keyword-based retrieval models such as BM25 may be sufficient to serve users. 

However, in reality, a user’s query is far from ideal especially because the user often does not have enough knowledge about the content of the relevant documents (i.e., a vocabulary gap). In such a case, a user’s query would not be effective. In some cases, the user might not be able to see any relevant result in the first page of search results. 

The current search engines do not provide much support for a user to reformulate a query. The goal of this task is to study how we can use algorithms to automatically reformulate a query to make it more effective, closer to an ideal query. 

\subsection{The Task}
\label{sec:search-task}
The search task focused on the challenge of query reformulation to bridge the vocabulary and semantic gaps inherent in task-oriented product search. Specifically, we gave participants a set of queries (described in \ref{sec:data-search}) asked them to submit reformulated queries in two distinct task variations: 
\begin{itemize}
    \item \textbf{Automatic Reformulation:} Teams submitted exactly one reformulated query for each original query. 
    \item \textbf{Interactive Reformulation:} Teams submitted up to four different reformulated versions of each original query. This simulated a user interface where a consumer could select the most relevant option. For final evaluation, performance was determined by the single best performing query from each of the reformulated queries. 
\end{itemize}
\subsection{Data and Queries}\label{sec:data-search}
We used the TREC 2024 Product Collection (see \ref{sec:data-over}) as our retrieval dataset. Participants were also given access to a standard BM25-based Search API to test their reformulations locally. Training and development datasets included original task-oriented ``complex'' queries, human-annotated reformulations, relevance judgments, and a subset of particularly difficult queries from the TREC 2024 test set where baseline BM25 scores were low.

Our test queries consisted of two main parts. The latter half of the queries (i.e. queries 051-100) were particularly difficult queries from the TREC 2024 test set as measured by their recall at 10 and 100. The other half of the queries (i.e. queries 001-050) were synthetic \textit{complex queries} generated following the following multi-step pipeline:
\begin{enumerate}
    \item \textbf{Complex Query Generation:} Product categories were fed into an LLM to generate complex, task-oriented queries formed by combining different product types both within and across those categories. For example, the LLM might combine product types related to office supplies to create the complex query: ``Home office makeover''
    \item \textbf{Deconstruction:} For each generated complex query, the LLM broke the overarching task down into its constituent parts to produce a series of simpler, foundational queries. For example, given the prior query, the LLM might create some queries such as: office chair, office desk, or filing cabinet.
    \item \textbf{Baseline Retrieval:} Each simple query was executed through the BM25 search API to retrieve the top 5 product results.
    \item \textbf{LLM-as-a-Judge Filtering:} An LLM was utilized as a judge to evaluate the top 5 retrieved results for relevance. If a relevant product was identified, the simple query qualified as viable.
\end{enumerate}
Ultimately, complex queries were selected for the final test collection only if at least $50\%$ of their constituent simple queries successfully qualified during the filtering process. 

We call these task-oriented queries \textit{complex queries} because they represent multiple search intents simultaneously. Moreover, these search intents are not necessarily directly apparent from the query. For example, if a user searches the query ``Home office makeover'' this implies a need for many product types (e.g. a office chair, a desk, and various supplies such as papers and pens). None of these product types are directly listed in the query, hence, to achieve good performance using BM25 these complex queries require reformulation. 

\subsection{Evaluation and Annotation}
Evaluation for the search task used the following metrics: 
\begin{itemize}
    \item \textbf{Task Completion NDCG:} Normalized Discounted Cumulative Gain calculated with an emphasis on retrieving all essential products for task completion
    \item \textbf{Essential Product Recall@K:} The proportion of essential products that appear in the top K results
    \item \textbf{Product Coverage Score:} A measure of how well the retrieved products cover the different aspects or requirements of the task
    \item \textbf{Average Precision:} The precision averaged over different recall points
    \item \textbf{Diversity Metrics:} Measures how well the retrieved results cover different product categories needed for the task    
\end{itemize}
For the Interactive Reformulation track, the best-performing reformulated query (according to the Task Completion NDCG) will be used for the final evaluation.

Products retrieved based on the submitted reformulated queries were pooled and assessed by human evaluators. Each product was be assigned a relevance score using the following criteria:
\begin{enumerate}
    \item \textbf{Essential (3):} The product is absolutely necessary to complete the task described in the original query
    \item \textbf{Highly Relevant (2):} The product is very useful for the task but might be substituted or is not strictly required
    \item \textbf{Somewhat Relevant (1):} The product has some utility for the task but is secondary rather than necessary
    \item \textbf{Not Relevant (0):} The product has no clear utility for the specific task described
\end{enumerate}

\subsection{Results and Discussion}
\label{sec:search-results}

\begin{figure}
    \centering
    \includegraphics[width=0.65\linewidth]{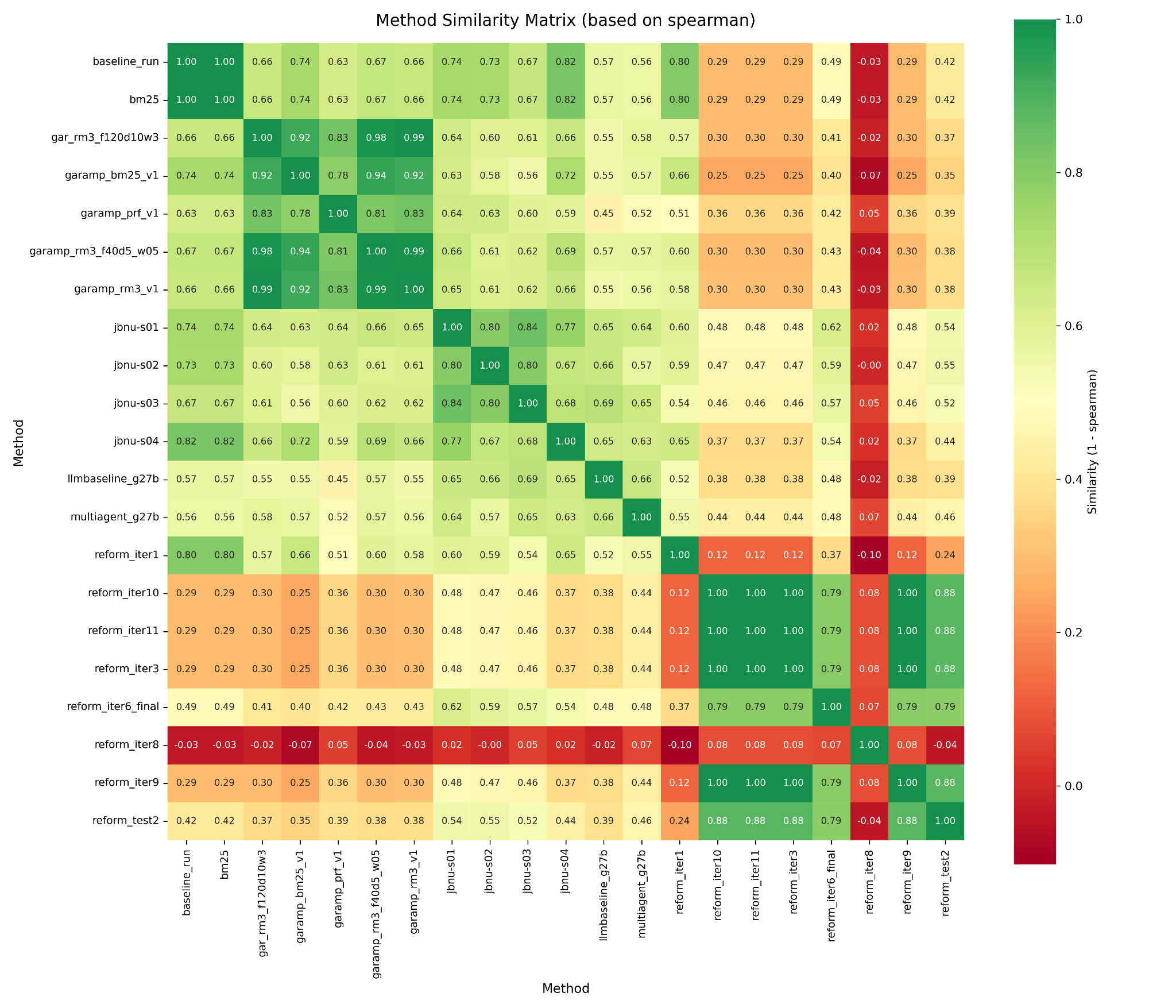}
    \caption{Pairwise run similarity matrix, computed via Spearman's $\rho$ over per-query score vectors. Highly correlated clusters emerge for the baseline systems, the successive \texttt{reform\_iter} runs, and the participant RM3 expansion family. The \texttt{reform\_iter8} experiment remains isolated, exhibiting minimal or slight negative correlation with all other runs.}
    \label{fig:sim}
\end{figure}

\begin{figure}
    \centering
    \includegraphics[width=0.65\linewidth]{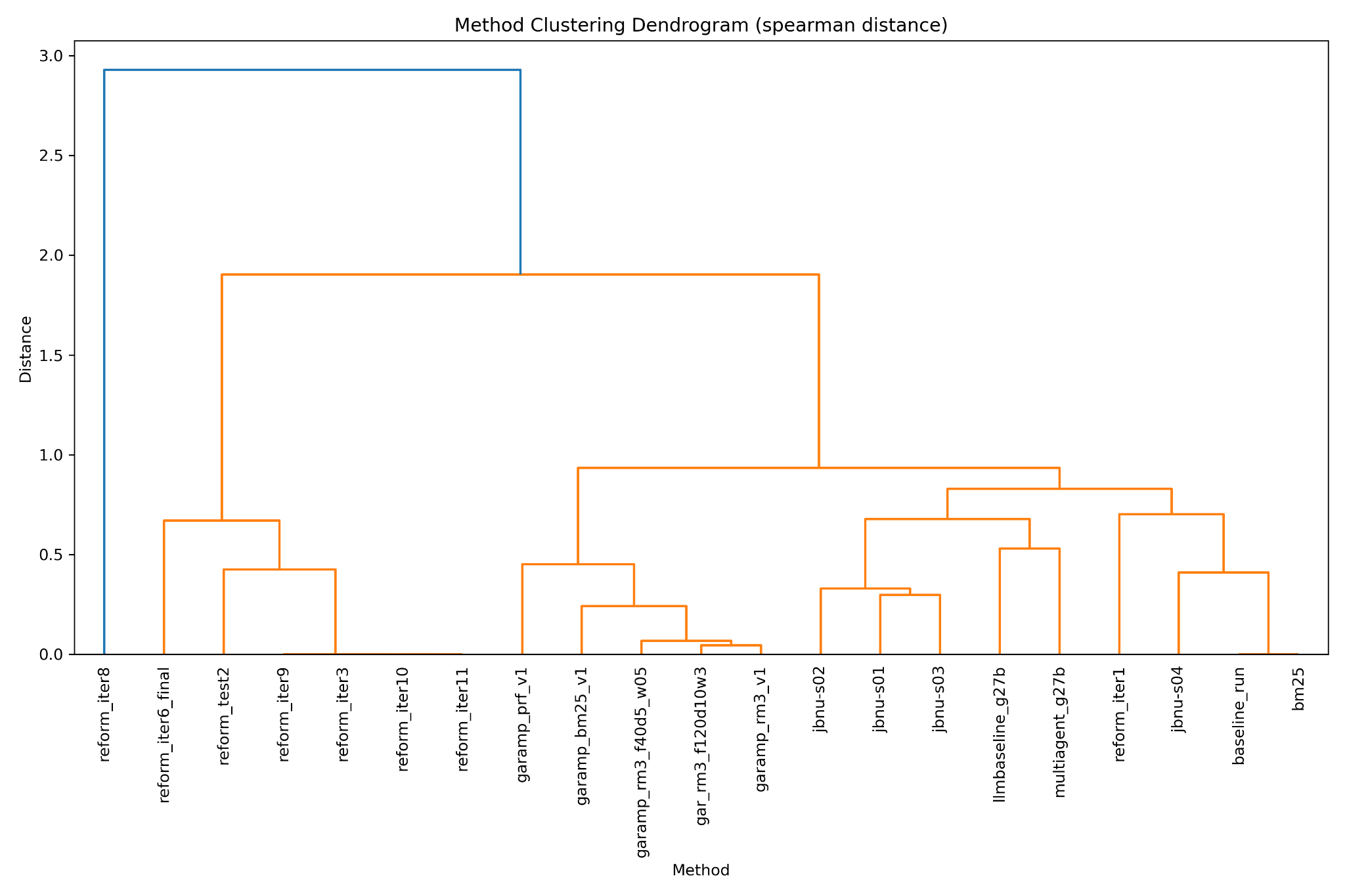}
    \caption{Agglomerative hierarchical clustering of runs based on Spearman distance ($1-\rho$) over per-query scores. The analysis reveals three primary behavioral paradigms: the participant RM3 expansion family; the organizer's iterative LLM loops; and a heterogeneous cluster combining the \texttt{jbnu} submissions, the single/multi-agent generative models, and the baseline BM25 query.}
    \label{fig:dendro}
\end{figure}

We evaluate all retrieval strategies against the unmodified BM25 query baseline, designated as the organizer reference \texttt{bm25} (experimentally verified as identical to \texttt{baseline\_run} with a rank correlation of $1.00$, as shown in \ref{fig:sim},\ref{fig:dendro}). The evaluation cohort has two primary participant submissions: the \texttt{gar}/\texttt{garamp} run family and the \texttt{jbnu-s01}--\texttt{s04} runs. Additionally, it features the official LLM reformulation baseline , provided to participants (\texttt{llmbaseline\_g27b}) and a suite of informal organizer exploratory experiments (\texttt{reform\_*}, \texttt{multiagent\_g27b}). 

Performance is disaggregated across two evaluation strata defined in \ref{sec:data-search}: the \emph{Hard} stratum (comprising TREC~2024 queries selected due to poor baseline BM25 recall) and the synthetic \emph{Complex} stratum. The runs are evaluated using Mean Average Precision (MAP), task-completion nDCG, and essential-product recall at cutoffs $10$ and $100$. Crucially, task-completion nDCG utilizes the specific gain mapping induced by our annotation schema: Essential products ($=3$) are weighted at $10$, Highly Relevant ($=2$) at $1$, and Somewhat or Not Relevant items at $0$. This metric prioritizes the retrieval of product configurations strictly required to complete a multi-step task over simple topical relevance matches.

\begin{figure}
    \centering
    \includegraphics[width=0.7\linewidth]{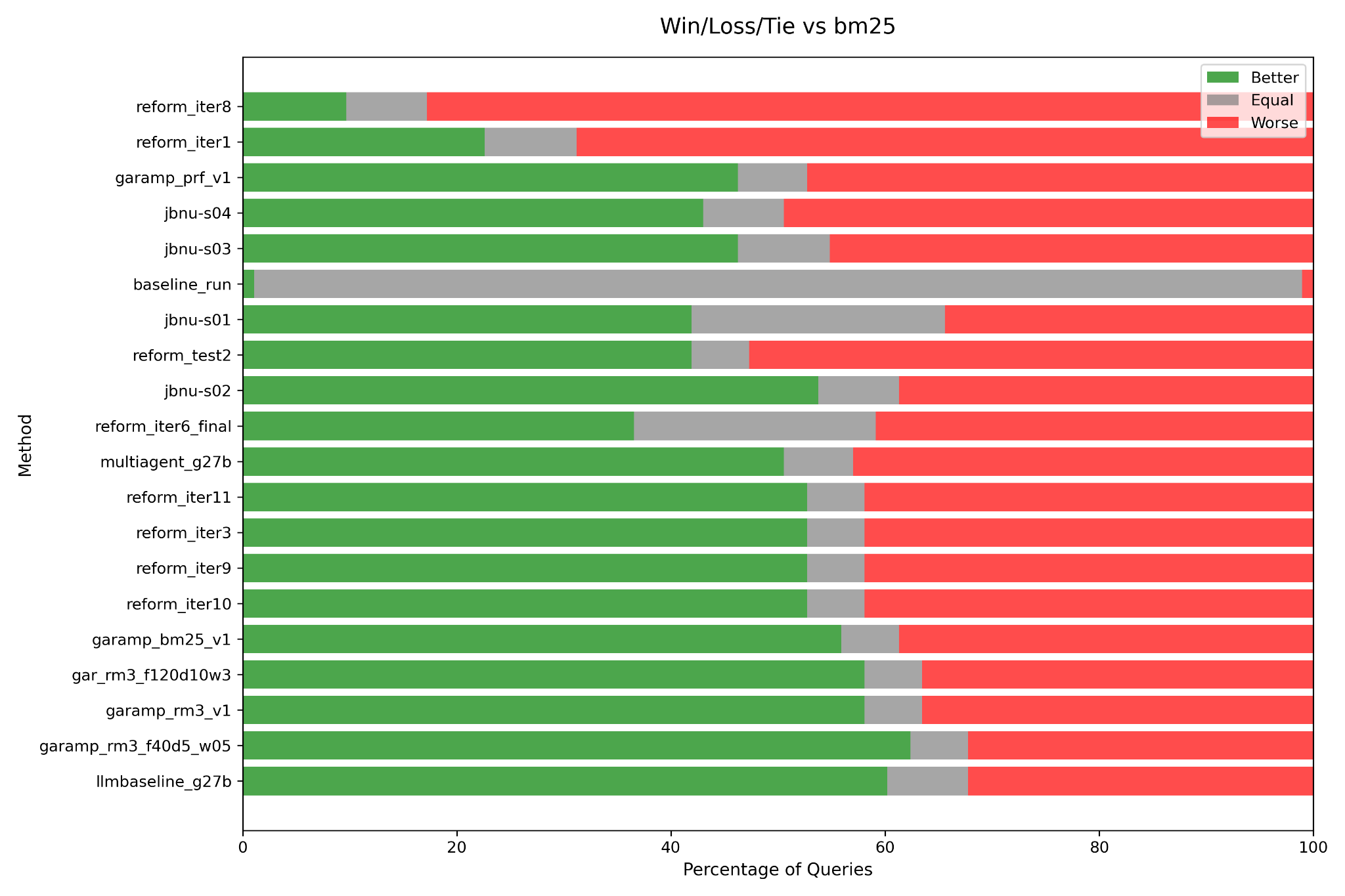}
    \caption{Mean per-query performance delta relative to the BM25 reference, averaged across all metrics and both evaluation strata. Positive values (green) denote net improvement over the baseline query; negative values (red) indicate degradation. The most pronounced average gains are achieved by the organizer LLM baseline and the participant RM3 expansion runs, whereas the informal \texttt{reform\_iter8} run exhibits catastrophic degradation driven by failure on the Hard stratum.}
    \label{fig:delta}
\end{figure}

\begin{figure}
    \centering
    \includegraphics[width=0.7\linewidth]{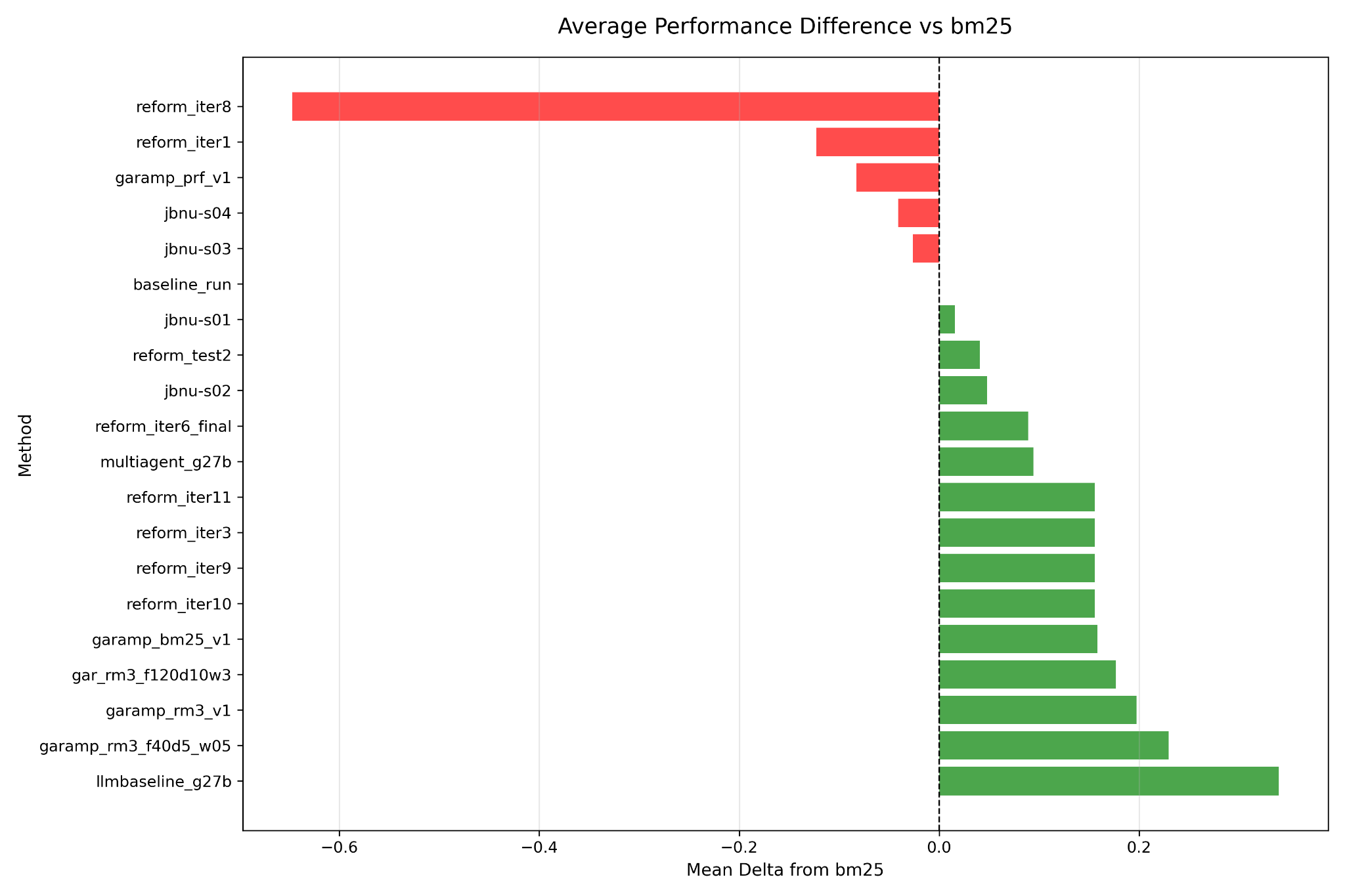}
    \caption{Per-query outcome distribution against the BM25 reference, plotting the fraction of queries where a given run improves, equals, or degrades performance relative to \texttt{bm25} under task-completion nDCG. The complete uniformity of \texttt{baseline\_run} confirms its identity with \texttt{bm25}. No evaluated approach improves upon more than approximately $60\%$ of the query set, and even the top-performing methods exhibit a substantial degradation footprint.}
    \label{fig:winloss}
\end{figure}

\paragraph{Aggregate Performance and Reformulation Trade-offs.}
When aggregated across both strata and all evaluation metrics (\ref{fig:delta}), the highest mean gains over the BM25 baseline are generated by the organizer LLM baseline (\texttt{llmbaseline\_g27b}), closely followed by the participant RM3 expansion models (\texttt{garamp\_rm3\_f40d5\_w05}, \texttt{garamp\_rm3\_v1}, and \texttt{gar\_rm3\_f120d10w3}). Conversely, a subset of systems demonstrate net-negative average utility, notably the participant \texttt{garamp\_prf\_v1} system and the organizer's exploratory \texttt{reform\_iter1} and \texttt{reform\_iter8} runs. 

The per-query performance distribution (\ref{fig:winloss}) reveals an important divergence between mean margin and win frequency. While \texttt{llmbaseline\_g27b} yields the highest absolute mean delta, the participant \texttt{garamp\_rm3\_f40d5\_w05} run improves the largest outright proportion of individual queries ($\approx 62\%$). Across all high-performing runs, a consistent behavior emerges: while query reformulation yields net-positive utility on average, no single method consistently outperforms BM25 on a strict per-query basis. This variance validates the ideal-query premise introduced in \ref{sec:search-task}; highly effective reformulations exist within the solution space for most information needs, but contemporary automated methods lack the robustness to consistently locate them.

\begin{figure}
    \centering
    \includegraphics[width=0.9\linewidth]{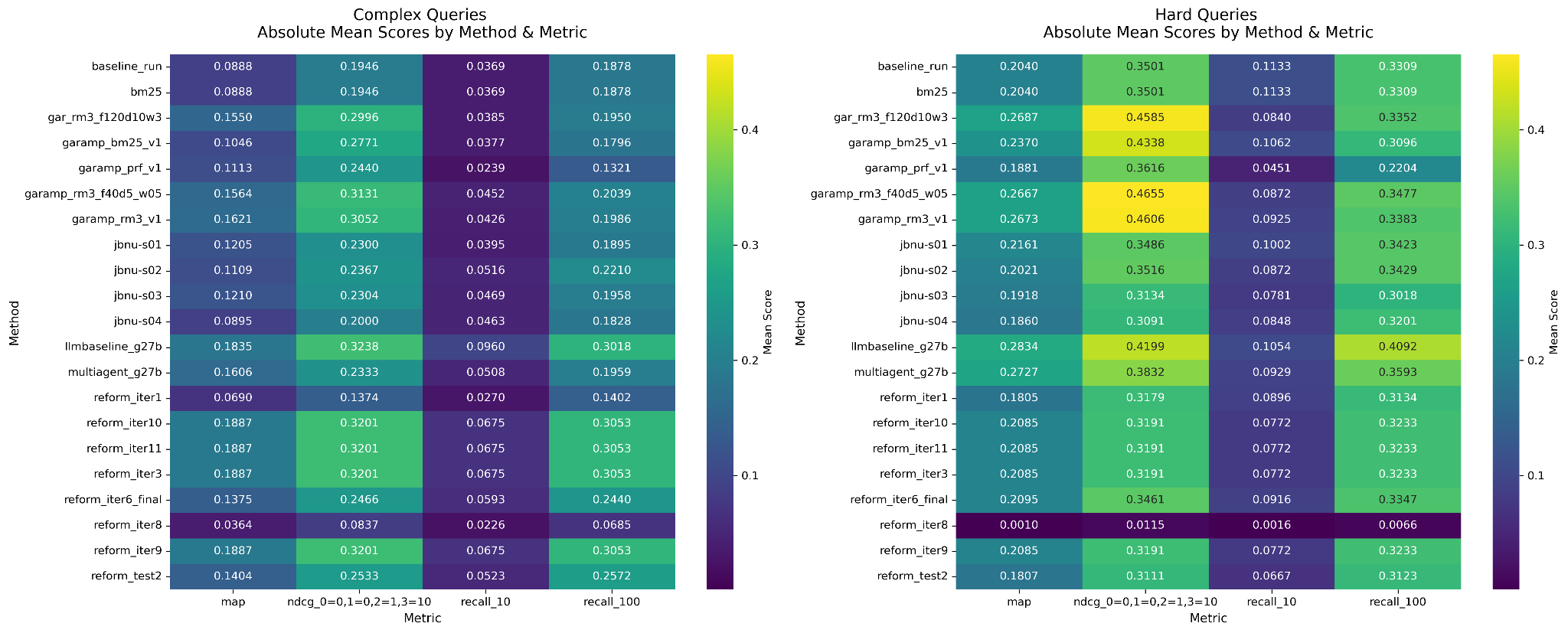}
    \caption{Absolute mean scores stratified by run (rows) and evaluation metric (columns) across the Complex (left) and Hard (right) strata. Absolute scores are higher on the Hard stratum despite its selection for low baseline recall, illustrating that the two evaluation subsets represent structurally distinct failure modes for lexical search engines (see text).}
    \label{fig:heatmaps}
\end{figure}

\paragraph{Stratum-Specific Evaluation.}
The stratified results in \ref{fig:heatmaps} demonstrate that aggregate rankings obscure critical behavioral differences between the two evaluation subsets. On the \emph{Hard} query stratum, the participant RM3 expansion runs achieve the highest task-completion nDCG scores ($0.459$--$0.466$), outperforming the organizer LLM baseline ($0.420$), while achieving comparable MAP performance ($\approx 0.267$ vs.\ $0.283$). The advantage of the LLM baseline on this stratum is isolated to deep recall at cutoff 100 ($0.409$ vs.\ $\approx 0.34$). This indicates two distinct mechanisms for overcoming the baseline vocabulary gap: classical pseudo-relevance feedback (PRF) methods optimize the precision ranking of essential products, whereas the generative LLM baseline maximizes absolute recall. Notably, neither paradigm improves early precision on this stratum; the baseline BM25 query maintains the highest recall@10 ($0.113$) across all tested systems. For lexical-gap queries, reformulation serves to reorder and deepen the lower tiers of the ranking rather than elevating essential products into the top ten results.

The \emph{Complex} stratum presents an inverted performance profile. While absolute scores are suppressed across all systems, reformulation successfully early precision. Here, \texttt{llmbaseline\_g27b} achieves a recall@10 of $0.096$ compared to the baseline BM25 score of $0.037$, and the exploratory \texttt{reform\_iter} runs remain highly competitive on MAP and recall@100. Consequently, automated reformulation acts as a mechanism to surface essential products early for complex compositional needs, whereas it primarily functions as a deep reordering mechanism for standard lexical-gap queries.

\begin{figure}
    \centering
    \includegraphics[width=0.9\linewidth]{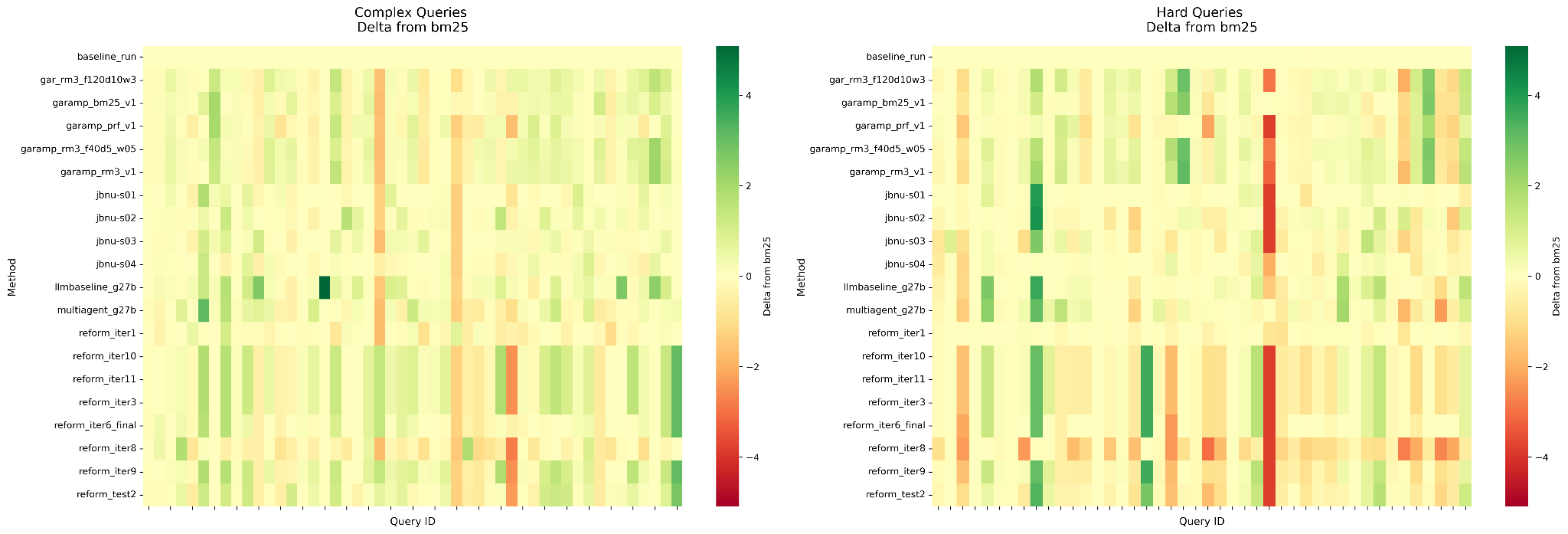}
    \caption{Per-query delta relative to BM25, mapped by run (rows) and individual query (columns) for the Complex (left) and Hard (right) strata. Green cells denote performance gains; red cells denote degradation. Variance is strongly concentrated within specific query columns, illustrating that reformulation utility is primarily dictated by query characteristics rather than algorithmic variations.}
    \label{fig:perquery}
\end{figure}

\paragraph{Query-Level Sensitivity Analysis.}
The per-query delta visualizations (\ref{fig:perquery}) contextualize the variance observed in the win/loss distributions. The vast majority of query-run pairs exhibit deltas near zero, indicating that aggregate performance shifts are heavily driven by a distinct minority of queries. Furthermore, this variance is highly structured and columnar: specific queries systematically show positive or negative deltas across all independent systems regardless of the underlying algorithm. 

This behavior indicates that the success of query reformulation is primarily an intrinsic property of the query itself rather than the optimization method. A small subset of queries proves uniformly hostile to reformulation, appearing as persistent dark-red columns on the Hard stratum where every system underperforms BM25. These systematically difficult queries bound the maximum achievable aggregate gain and explain the persistent degradation rates among even the top-performing systems. These findings suggest that implementing query-level routing mechanisms represents a optimization lever at least as critical as the refinement of the reformulation algorithms themselves.

\paragraph{System Behavioral Clustering.}
The similarity matrix and corresponding dendrogram (\ref{fig:sim},\ref{fig:dendro}) categorize the evaluated systems by their rank-ordering behavior rather than absolute performance metrics. The participant RM3 expansion runs converge into a highly tightly coupled cluster. 

Another distinct cluster groups the \texttt{jbnu} submissions, the single- and multi-agent LLM architectures, and the baseline BM25 reference system. This clustering indicates that these generative reformulations introduce more conservative perturbations to the original document ranking compared to the aggressive expansions of the PRF models. 
\section{Recommendations Task}
\label{sec:recommendation}


\begin{figure}[h]
    \centering
    \begin{subfigure}[b]{0.48\linewidth}
        \centering
        \includegraphics[width=\linewidth]{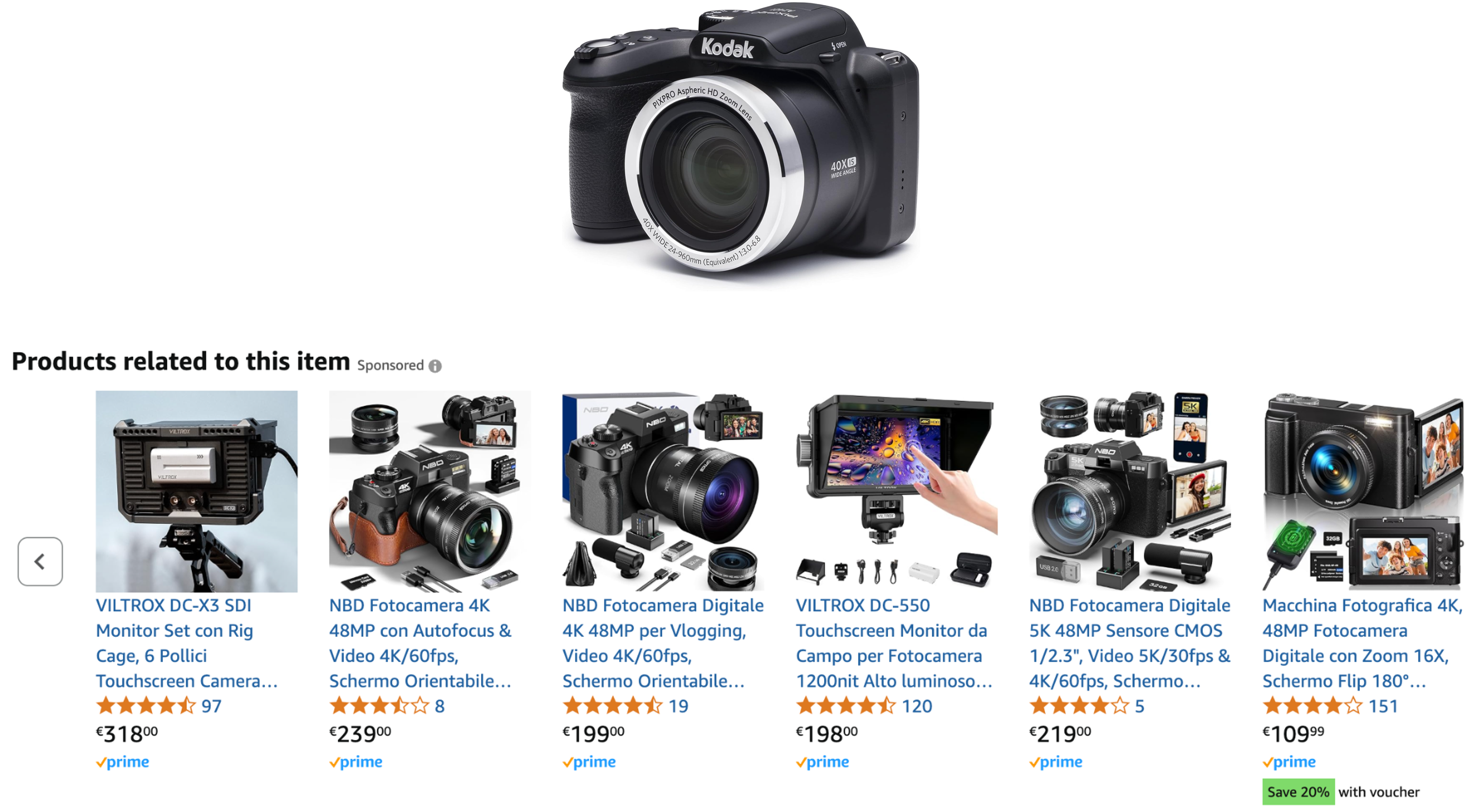}
    \end{subfigure}
    \hfill
    \begin{subfigure}[b]{0.48\linewidth}
        \centering
        \includegraphics[width=\linewidth]{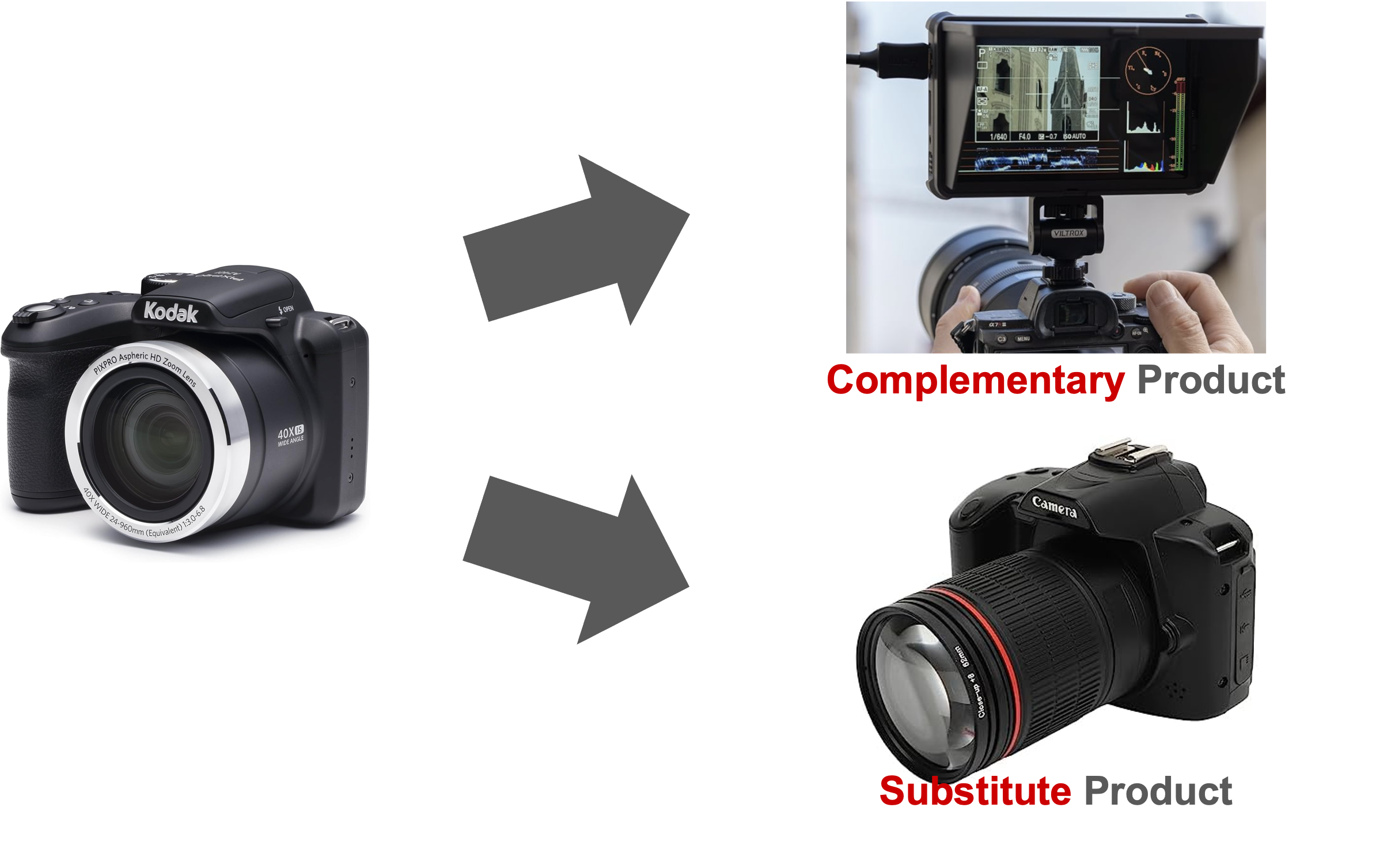}
    \end{subfigure}
    
    \caption{Left: A typical e-commerce related products carousel showing a mix of item relationships. Right: Our task focuses on explicitly distinguishing between complementary and substitute products.}
    \label{fig:combined_rec}
\end{figure}

In the recommendations task, we focus on item-to-item relationships. One example of how this could be used is around helping users find related products of \textit{different kinds} --- there are several different dimensions or types of relationships that can cause a user to to purchase a product or not, and in particular to co-purchase a pair of products or not. These include:

\begin{itemize}
    \item \textbf{Complementary products} are products that function together, such as gaming consoles and televisions.  If a user has bought a product, they may be interested in its complements.
    \item \textbf{Substitute products} are products that functionally substitute for each other, such as different models of television.  If a user has already bought one such product, they probably do not want its substitutes; however, if they are in the process of selecting a product, they may be very interested in substitutes.
    \item Products may or may not be \textbf{compatible} with each other, or with the user's needs or current products. For example, ``phone case'' and ``phone'' are complementary product categories, but an Android case is incompatible with an iPhone.
    \item Products may or may not align with the user's \textbf{preference}.  This is similar to compatibility, except that it is fully subjective to the user's preferences rather than an objective relationship that could be applied to known information about the user (either their current products, or information about them such as shoe size).
\end{itemize}

To date, there are no good data sets for training and evaluating models capable of discovering such nuanced product relationships. Existing definitions used to derive such relationships from data are also often unsatisfying (e.g. defining complementarity in terms of simultaneous purchase for a ``joint information need''~\citep{zhang2018complementqualityaware, liu2020decoupledsubstitutecomplement, papso2023longtailcomplementary}, ruling out purchasing complementary products in separate sessions).
We take a more functional definition: complementary and substitutability are defined in terms of whether the products perform the same function, or whether they work together performing different functions.

Building product relationship models would enable compelling new product search and recommendation applications: systems can provide recommendations that are deeply connected to the user's particular task, such as substitutes when they are looking to understand the options to meet a need, or a set of products that are complementary to the user's last purchase but substitutes for each other when the user returns to the system after making a significant purchase.  It will also enable new types of recommendation explanation that involve a recommended item's relationship to other items in a slate (or excluded from it), the user's current and past purchases, and other item-related contextual information.

\subsection{The Task}
Queries consisted of Amazon product IDs. Each team submitted three separate rankings for each product:
\begin{itemize}
    \item Top 10 complementary products.
    \item Top 10 substitute products.
    \item Top 100 related products, complete with substitute/complement labels (referred to as a ``pool run''), with the goal of developing deeper pools.
\end{itemize}

\subsection{Evaluation Methodology}
Relevance grades are provided by NIST assessors. The evaluation metrics are defined as follows:
\begin{itemize}
    \item NDCG calculated separately on the complement and substitute lists.
    \item NDCG evaluated on the pooled lists, assigning 1/2 credit for items that are ranked but mislabeled.
    \item Agreement (measured via Cohen's $\kappa$) between the pooled labels and the assessors.
    \item Diversity (measured via intra-list similarity) for the complement and substitute lists.
\end{itemize}

\subsection{Results and Discussion}

In the evaluation phase, a total of 7634 product/query pairs were annotated by the NIST assessors. An analysis of these annotations revealed that the majority of the pairs, specifically 54.2\%, were deemed non-relevant. Furthermore, 15.1\% of the pairs were marked as ``unable-to-assess''. Because only a single team submitted runs for this task, the resulting pool of judged documents is limited, which presents a challenge for conducting extensive future work using this specific data pool.

\begin{figure}[t]
    \centering
    \includegraphics[width=0.7\linewidth]{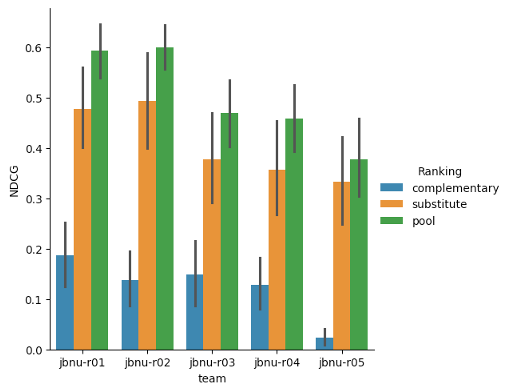}
    \caption{NDCG performance of the participant's runs (jbnu-r01 to jbnu-r05) across complementary, substitute, and pool rankings. The results indicate that retrieving complementary items is consistently more challenging.}
    \label{fig:ndcg_results}
\end{figure}

The performance of the submitted runs (labeled jbnu-r01 through jbnu-r05) was evaluated using the NDCG metric across three distinct ranking categories: complementary, substitute, and a combined pool. Figure \ref{fig:ndcg_results} illustrates the NDCG scores achieved by each run. A clear trend across all submissions is that the models performed significantly better on the substitute and pool tasks compared to the complementary task. For instance, across all runs, the NDCG for substitute items was consistently higher than that for complementary items. This performance gap highlights a critical insight: the task of identifying complementary products appears to be inherently more difficult than identifying substitute products. 

We hope this data would be reusable in various tasks, e.g., complementary product recommendation, substitute product recommendation, and basket-completion. For example, when evaluating a product recommendation, even if the product is not a perfect match, it can be given partial credit if it was a substitute.

Looking forward, the results and limitations of this year's track bring up several open questions for the community. First, there is a need to explore how we can create a more robust collection of realistic user tasks or needs, potentially by leveraging user simulation techniques. Second, researchers must consider how to effectively support conversational search and recommendation paradigms through a unified matching model. Finally, our ultimate goal should be to move beyond standard search and recommendation outputs to provide comprehensive decision support for users.
\section{Conclusion}
\label{sec:conclusion}
Despite the widespread usage of search engines in e-commerce, there is no high-quality dataset designed to evaluate end-to-end retrieval quality. In 2025, we ran a revised and continued version of the Product Search track previously run at TREC 2023 and TREC 2024 focusing on two tasks: \textbf{query expansion} and \textbf{related-product recommendation}. In this paper we have described each task in detail and have provided some initial analysis of results. We anticipate the data from this track will enable better recommendation and search applications that reflect user needs, as a building block for conversational product discovery experiences.

\bibliography{anthology,custom}
\bibliographystyle{chicago}
\end{document}